\documentstyle[graphicx,float]{article}

\begin{document}
\title{The Bragg regime of the two-particle Kapitza-Dirac effect}
\author{Pedro Sancho \\ Centro de L\'aseres Pulsados, CLPU, E-37008, Salamanca, Spain}
\date{}
\maketitle
\begin{abstract}
We analyze the Bragg regime of the two-particle Kapitza-Dirac
arrangement, completing the basic theory of this effect. We provide
a detailed evaluation of the detection probabilities for multi-mode
states, showing that a complete description must include the
interaction time in addition to the usual dimensionless parameter
$w$. The arrangement can be used as a massive two-particle beam
splitter. In this respect, we present a comparison with
Hong-Ou-Mandel-type experiments in quantum optics. The analysis
reveals the presence of dips for massive bosons and a differentiated
behavior of distinguishable and identical particles in an unexplored
scenario. We suggest that the arrangement can provide the basis for
symmetrization verification schemes.
\end{abstract}

\vspace{4mm}


\vspace{4mm}

\section{Introduction}

The Kapitza-Dirac proposal \cite{KD} provides a beautiful
demonstration of the diffraction of massive particles by standing
light waves. The proposal has been realized experimentally for atoms
and electrons \cite{Gou,Mar,Cah,Ba1,Ba2}.

More recently, it has been suggested that additional effects could
be present when we move from one- to two-particle massive systems
interacting with the optical diffraction grating \cite{san}. In this
type of arrangement two different dynamics take place
simultaneously. On the one hand, the particles interact with the
optical wave generating diffraction patterns. On the other hand, if
the two particles are identical, the exchange effects must also be
taken into account. The resultant joint dynamics shows a much richer
behavior.

As it is well-known \cite{HBa}, there are two regimes in the
Kapitza-Dirac effect, diffraction and Bragg scattering. In
\cite{san} the first one was studied for two-particle systems. Here,
we complete the basic analysis of the two-particle Kapitza-Dirac effect by
considering two-particle Bragg scattering.

As we did in \cite{san}, we consider separately single- and
multi-mode states. In contrast with \cite{san}, where we only gave a
qualitative description of the second ones, we present here a simple
model of the problem that allows for a detailed quantitative
evaluation of the detection probabilities in the multi-mode case.
Our model takes into account the dependence of the width of the
window of modes that can be scattered on the duration of the
interaction \cite{Pr1,Pri}. Because of this dependence, the
probabilities of transmission and reflection of multi-mode states
must be expressed in terms of the interaction time, in addition to
the dimensionless parameter $w$, which is the only parameter present
in the case of single-mode states.

Although the Bragg scattering of massive particles has been
extensively studied, specially in BEC, there are some aspects of the
two-particle problem that still deserve attention. In particular,
the behavior of massive particles in Hong-Ou-Mandel (HOM)-type
experiments \cite{HOM} remains rather unexplored. As these
experiments have played an important role in quantum optics it seems
necessary to analyze its massive counterpart. In this respect, it
has been many times suggested in the literature the possibility of
using the Bragg regime of the Kapitza-Dirac  effect as a basis for
massive beam splitters \cite{Pri,Ada}. This is a natural choice
because it generates two possible exit paths for the particles, just
as a beam splitter. As a simple extension of these ideas, it is
natural to think of the two-particle Bragg scattering as a serious
candidate for the implementation of massive two-particle beam
splitters.

Several results emerge from our analysis: (i) We show that, just as
for photons, one can observe dips in the case of massive bosons.
(ii) We have, as in quantum optics \cite{Lou,Scu}, that the behavior
of distinguishable and identical particles is the same if the
parallel momenta of the two particles are equal. However, if the two
particles are in different multi-mode states we can observe
different behaviors. This is a previously not considered scenario,
which deserves attention. (iii) Finally, we shall propose a possible
application of the arrangement. The two-particle Bragg scattering
could be used to test the (anti)symmetrization of the wave functions
of pairs of identical particles.

The plan of the paper is as follows. In Sect. 2 we briefly review
the fundamentals of the Bragg regime in the Kapitza-Dirac effect, and
we discuss the different situations present in its two-particle
extension. We devote Sects. 3 and 4 to the evaluation of the
detection probabilities for, respectively, single- and multi-mode
states. The possibility of using the two-particle Kapitza-Dirac
arrangement as a massive two-particle beam splitter is presented in
the Discussion where, in addition to recapitulate on the main
results of the paper, we compare our approach with HOM-type
experiments.

\section{General considerations}

The Bragg scattering is the relevant process for thick standing
waves with weak associated potentials. When these two conditions are
fulfilled the diffraction can only take place for some particular
angles, the Bragg angles. This behavior contrasts with that observed
for thin waves, where diffraction occurs for any angle of incidence
and many different diffraction orders can be reached. On the other
hand, if the potential is strong, we have coherent channeling.

The theory of one-particle Bragg scattering by a standing light wave
can be found in \cite{HBa}. In this reference there is also an
excellent discussion of the physical and mathematical differences
between the Bragg and diffraction regimes. We denote by $k_L$ the
wave number of the optical grating, usually a laser beam. When the
particle is incident on the grating at the Bragg angle, $\hbar
k_L/p$ with $p$ the total momentum of the particle, the energy and
momentum are conserved. In addition, the particles incident exactly
at the first order of the grating (the momentum parallel to the
grating equal to $\hbar k_L$) can be scattered into the $-1st$ order
with a momentum change $2\hbar k_L$. Moreover, the transitions to
other orders are forbidden.

From a more mathematical point of view, the (first-order) Bragg
angle $\theta $ is given by the expression $\lambda =d_L \sin \theta
$, where $\lambda $ is the de Broglie wavelength of the particles
and $d_L=\lambda _L/2$ is the periodicity of the light beam. The
interaction of the particle with the grating is ruled by the
potential $V=V_0 \cos ^2 k_Lx$, with $x$ denoting the coordinate
parallel to the grating. As usual, the solution of the quantum
equation of evolution is obtained by introducing wave functions of
the form $\psi (x,X)= \sum _n c_n \exp (i(nk_Lx+KX))$, with $X$ the
coordinate perpendicular to the grating and $K$ the initial wave
number in that axis (which does not change because the interaction
along it is null). In the Bragg regime, being the incident wave
function in the state $n=1$, the final state of the particle can
only be $n=\pm 1$ with coefficients \cite{HBa}:
\begin{equation}
c_+ = e^{-i\epsilon \tau } \cos w \;
; \; c_- = -ie^{-i\epsilon \tau} \sin w \; ; \; w = \frac{V_0 \tau }{4\hbar}
\end{equation}
where $\tau $ denotes the duration of the interaction, $\epsilon =\hbar k_L^2/2m$, and $c_+=c_{1}$ and
$c_-=c_{-1}$. The above equation shows an oscillatory behavior of
the probabilities of finding the particle in each of the orders
$n=\pm 1$ as a function of $V_0\tau $.

Note that although the incident $x$-component of the momentum of the
particle is fixed to $\hbar k_L$, varying $K$ we can have different
Bragg's angles.

When we have two incident particles we have a wider range of
possibilities. They  are depicted in Fig. 1:
\begin{figure}[H]
\center
\includegraphics[width=9cm,height=7cm]{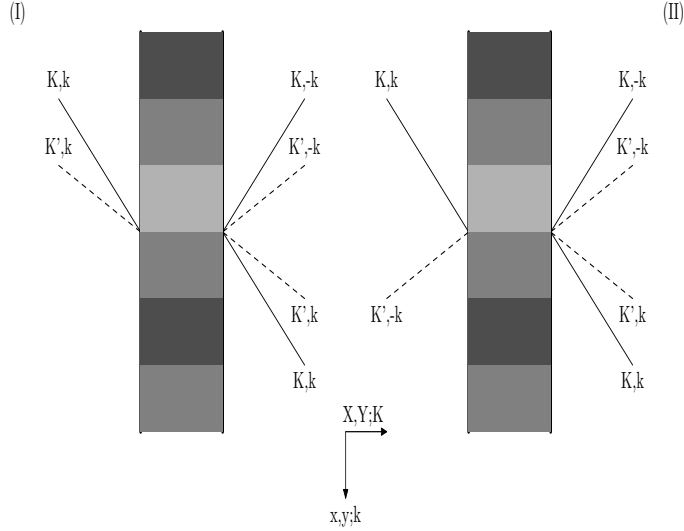}
\caption{Two particles incident on a standing light wave. The full
and dashed lines represent, respectively, particles with
momenta perpendicular to the light wave $K$ and $K'$. The parallel momenta
$\pm k$ are $\pm k_L$ (or values very close to them, see the
multi-mode section).}
\end{figure}
The part $(I)$ of the figure corresponds to the case in which both
particles arrive to the grating with the same momentum in the
direction parallel to the light beam, $k_L$ (in the multi-mode realm
we must also consider values of $k$ very close but not equal to
$k_L$, see Sect. 4). As signaled before, taking different momenta in
the perpendicular axis we can have different Bragg's angles (the
angle between the path and the normal to the grating). On the other
hand, the part $(II)$ of the figure represents the situation where
the parallel momenta of the two particles are opposite.

In the case $(I)$ the wave functions of the two particles after the interaction are
\begin{equation}
\psi _{k_L,K} (x,X)=c_+ e^{i(k_Lx+KX)} + c_- e^{i(-k_Lx+KX)}
\end{equation}
and
\begin{equation}
\psi _{k_L,K'} (y,Y)=c_+ e^{i(k_Ly+K'Y)} + c_- e^{i(-k_Ly+K'Y)}
\end{equation}
where $y$ and $Y$ denote the coordinates of the other particle.

From these expressions one can derive, as in \cite{san}, the wave
functions in momentum space. Instead, we move to the more concise
brackets formalism, where we have $|k_L,K>_1=c_+ |k_L>_1|K>_1 +
c_-|-k_L>_1|K>_1$ and $|k_L,K'>_2=c_+ |k_L>_2|K'>_2 +
c_-|-k_L>_2|K'>_2$, with the subscripts $1$ and $2$ denoting the two
particles.

When the two particles are distinguishable the complete states are
$|I> _{dis} = |k_L,K>_1|k_L,K' >_2$ and $|II> _{dis} =
|k_L,K>_1|-k_L,K' >_2$ (where now $|-k_L,K'>_2=c_+ |-k_L>_2|K'>_2 +
c_-|k_L>_2|K'>_2$). In contrast, if the particles are identical the
wave function must be (anti)symmetrized, that is, $|I> =N |k_L,K>_1
|k_L,K' >_2 \pm N|k_L,K'>_1|k_L,K >_2$ and $|II> =N |k_L,K>_1
|-k_L,K' >_2 \pm N|-k_L,K'>_1|k_L,K >_2$, where the upper sign holds
for bosons and the lower one for fermions. $N$ is the normalization
coefficient, which will be determined later.

We assume that the particles are in the same (or symmetric) spin and
electronic states. Thus, as done above, the spatial part of the wave
function must be symmetric for bosons and antisymmetric for
fermions. By simplicity, the spin and electronic variables can be
dropped from all the expressions. The extension to antisymmetric
spin or electronic states is straightforward.

\section{Single-mode states}

This section is devoted to the simple case of single-mode states.
With this simplification it is easy to illustrate the properties
of the system. Later, in the next section, we move to the more
realistic case of multi-mode states.

\subsection{Distinguishable particles}

In order to later compare with identical particles, we
assume both distinguishable particles to be characterized by the
same $c_{\pm}$ coefficients (both masses to be equal). Using the
explicit expression for $|I>_{dis}$, the probabilities are easily
evaluated
\begin{eqnarray}
{\cal P}_{dis}^{(I)}(k_L,K;k_L,K')= |c_+|^4 \; ; \; {\cal P}_{dis} ^{(I)}(k_L,K;-k_L,K')= |c_+|^2|c_-|^2   \nonumber \\
{\cal P}_{dis}^{(I)}(-k_L,K;-k_L,K')= |c_-|^4 \; ; \;
{\cal P}_{dis} ^{(I)}(-k_L,K; k_L,K')= |c_-|^2|c_+|^2
\end{eqnarray}
In a similar way, we obtain for the case $(II)$:
\begin{eqnarray}
{\cal P}_{dis}^{(II)}(k_L,K;-k_L,K')= |c_+|^4 \; ; \; {\cal P}_{dis} ^{(II)}(k_L,K;k_L,K')= |c_+|^2|c_-|^2  \nonumber \\
{\cal P}_{dis}^{(II)}(-k_L,K;k_L,K')= |c_-|^4 \; ; \; {\cal P}_{dis}
^{(II)}(-k_L,K; -k_L,K')= |c_-|^2|c_+|^2
\end{eqnarray}
Note that the probabilities for double transmission (${\cal
P}^{(I)}(k_L;k_L)$ and ${\cal P}^{(II)}(k_L;-k_L)$), one scattering (${\cal
P}^{(I)}(k_L;-k_L)$, ${\cal P}^{(I)}(-k_L;k_L)$, ${\cal P}^{(II)}(k_L;k_L)$ and
${\cal P}^{(II)}(-k_L;-k_L)$) and double scattering (${\cal
P}^{(I)}(-k_L;-k_L)$ and ${\cal P}^{(II)}(-k_L;k_L)$) are equal in both
cases. The sum of the four terms in each equation adds to one.

\subsection{Identical particles, case (I)}

As the initial state can be factored into their perpendicular and
parallel parts, a product form remains after the interaction:
\begin{eqnarray}
|I>=N(c_+|k_L>_1+c_-|-k_L>_1) \times \nonumber \\
(c_+|k_L>_2+c_-|-k_L>_2)(|K>_1|K'>_2 \pm |K'>_1 |K>_2)
\label{eq:ast}
\end{eqnarray}
The exchange effects correspond to the crossed terms in $<I|I>$. In
Eq. (\ref{eq:ast}) only the transversal part (capital variables) can
generate crossed effects. The squared modulus of the transversal
part gives $2\pm 2Re( _1<K|K'>_{12}<K'|K>_2)$. As $<K|K'>=\delta
(K-K')$ there are only exchange effects when $K=K'$. In more
physical terms, when $K \neq K'$ the two particles can be
distinguished and the probabilities for identical and
distinguishable particles are equal.  In the case $(I)$ there are no
exchange effects for fermions, because both should be in the
incident state ($k_L,K$), a preparation forbidden by Pauli's
exclusion principle.

Now we consider the case $K=K'$, only valid for bosons. Using also the
longitudinal part ({\it small letter variables}) of Eq.
(\ref{eq:ast}) we have
\begin{eqnarray}
{\cal P}^{(I)}(k_L;k_L)=4 N^2|c_+|^4   \; ; \; {\cal P}^{(I)}(-k_L;-k_L)= 4 N^2|c_-|^4 \nonumber \\
{\cal P}^{(I)}(k_L;-k_L) \equiv {\cal P}^{(I)}(k_L;-k_L) + {\cal
P}^{(I)}(-k_L;k_L) = 8N^2|c_+|^2|c_-|^2
\end{eqnarray}
Now we can determine the normalization of the state. This is done by
the condition that the sum of all the probabilities must be unit
$\sum _{i,j=\pm} {\cal P}^{(I)}(ik_L;jk_L)=1$. To use this condition
we assume that no particle is absorbed or deflected to other
momentum states; all the pairs of particles are detected in one of
the four above states. In other words, we restrict our
considerations to the postselected set in which the two particles
are detected in these output beams, and the problem can be described
by a pure state. From the former condition easily follows $N=1/2$.

Taking into account the normalization condition we see that the
probabilities for bosons and distinguishable particles (with $K=K'$)
are equal. We conclude that in the case $(I)$ the probabilities for
identical and distinguishable particles agree. Physically, this
result can be easily understood. For $K \neq K'$ the identical
particles can be distinguished. On the other hand, for $K = K'$
(only bosons) we have that the exchange term has the same form of
the direct terms. There is not a distinctive exchange effect,
because the two terms of the state are equal, $|I> \sim |k_L>_1|K>_1
|k_L>_2|K>_2 + |k_L>_1|K>_1 |k_L>_2|K>_2$. According to the standard
interpretation, different terms in the quantum state must represent
different alternatives for the system. However, in our case the two
alternatives are really the same, and the state reduces to that of
distinguishable particles. This result resembles that reported for
photons interacting at a beam splitter \cite{Lou,Scu} (see the
Discussion).

The probability distributions are represented in Fig. 2:
\begin{figure}[H]
\center
\includegraphics[width=10cm,height=7cm]{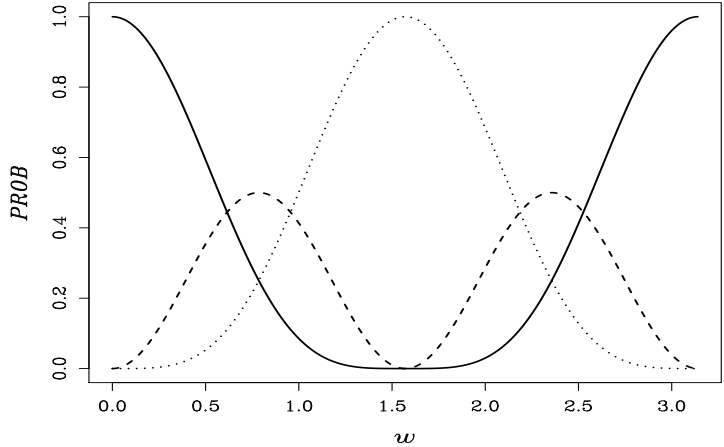}
\caption{Probabilities of bosons or distinguishable particles
detection (both are equal) for $K=K'$ in the scenario $(I)$ versus
$w$ (in arbitrary units). The continuous, dashed and dotted lines
represent respectively the cases ($k_L;k_L$), ($-k_L;-k_L$) and
($k_L;-k_L$).}
\end{figure}
A simple pattern can be observed. The probabilities of both
particles leaving the optical grating in the same parallel momentum
state and equal to the initial one $(k_L;k_L)$ shows an oscillatory
behavior.  The probability of observing one particle in each channel
$(k_L;-k_L)$ is also a periodic function, with large values when
those of the previous one are small. Finally, the probability of
having the two particles in the channel opposite to the initial one
is in general much smaller than the previous ones (except around the
crossing points $\cos ^4 w = \sin ^4 w$).

\subsection{Identical particles, case $(II)$}

The case $(II)$ is very similar. As in the case $(I)$ the particles
can be distinguished when $K \neq K'$. Then we concentrate on the
case $K=K'$. The most important difference between both situations
is that now we must also consider fermions, because the incident
particles are in different states (($k_L,K$) and ($-k_L,K$)) and
Pauli's exclusion principle does not forbid the preparation of that
state. The state after the interaction can be written as
\begin{eqnarray}
|II>_{K=K'}=N|K>_1|K>_2[(c_+^2 \pm c_-^2)(|k_L>_1|-k_L>_2 \pm |-k_L>_1|k_L>_2) + \nonumber \\ c_+c_- (1 \pm 1) (|k_L>_1|k_L>_2 + |-k_L>_1|-k_L>_2)]
\end{eqnarray}
The probabilities become
\begin{eqnarray}
{\cal P}^{(II)}(k_L;-k_L)/ N^2 \equiv ({\cal P}^{(II)}(k_L;-k_L) + {\cal
P}^{(II)}(-k_L;k_L))/ N^2 = \nonumber \\
2|c_+^2 \pm c_-^2|^2 = 2( |c_+|^2 \mp |c_-|^2)^2 \\
{\cal P}^{(II)}(k_L;k_L)/N^2=(1\pm 1)^2 |c_+|^2|c_-|^2 = {\cal
P}^{(II)}(-k_L; -k_L)/N^2 \nonumber
\end{eqnarray}
The normalization condition in this case is $N=1/\sqrt 2$.

The graphical representation is done in Fig. 3.
\begin{figure}[H]
\center
\includegraphics[width=10cm,height=7cm]{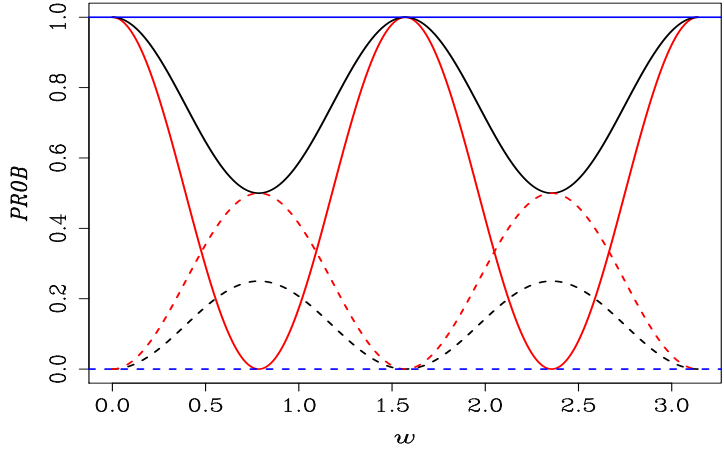}
\caption{The same as in Fig. 2 for the case $(II)$. The continuous
and dashed lines represent respectively the cases ($k_L;-k_L$) and
($k_L;k_L$) for distinguishable particles (black), bosons (red) and
fermions (blue). The curves for ($-k_L;-k_L$) are the same as for ($k_L;k_L$).}
\end{figure}
Now, for distinguishable particles the probability of leaving the
interaction region in different states is always larger than to do
it in any of the other channels. Initially, the two bosons are
in different states, but after passing through the optical
grating the probability of the two particles to be in the same state
reaches the $2\sin ^2 w \cos ^2 w$ value for any of the channels.
The probability of finding the outgoing bosons in the same state is
always larger than that for distinguishable particles. Moreover, for
some values of $w$, the probability of finding the two bosons in
different exit channels vanishes, giving rise to the presence of a
dip. This contrasts with the behavior of distinguishable particles,
for which that probability is never null. For fermions, the
probability of being the two in the same channel is forbidden by
Pauli's exclusion principle (also $K=K'$). The two fermions must
always be found in different channels.

\section{Multi-mode states}

We move now to the more realistic case of multi-mode states. For the
sake of clarity in the presentation we consider first a single particle.

\subsection{Single particle}

In order to have an analytically solvable model we assume both mode
distributions to be Gaussian functions before the interaction with the light grating:
\begin{equation}
f(k,K)=g_{k_0}(k)G_{K_0}(K)={\cal N}_g \exp (-(k-k_0)^2/\sigma ^2)
{\cal N}_G \exp (-(K-K_0)^2/\mu ^2)
\end{equation}
where the normalization factors ${\cal N}_g$ and ${\cal N}_G$ are
determined from the conditions $\int dk |g(k)|^2=1$ and $\int dK
|G(K)|^2=1$: ${\cal N}_g=(2/\pi \sigma ^2)^{1/4}, \cdots $. The
central value of the first distribution is $k_0 =\pm k_L$. We shall
only consider the case $k_0=k_L$, being the extension to $k_0=-k_L$
trivial.

After the interaction we consider first the {\it small letter
variables}. The Bragg scattering only takes place for modes whose
momenta are very close to the Bragg angle \cite{Ba2}. To be
concrete, the spread in velocity of particles that can be
diffracted, $\sigma _v$, depends on the interaction time with the
grating: $\sigma _v=1/(\tau k_L)$. This relation can be easily
derived from the time-energy uncertainty relation \cite{Pri}. This
property has been used to study the velocity distribution of BEC's
because the velocity selectivity of the previous condition
\cite{Pr1}. In terms of wave vectors, this condition can be
rewritten as $\sigma _k =m/(\tau \hbar k_L)$. Then the modes in the
interval $[k_L -\frac{1}{2} \sigma _k, k_L +\frac{1}{2}\sigma _k]$
can be scattered, whereas the modes obeying $|k-k_L|>\sigma _k/2$
are always transmitted without possibility of scattering. The
probability to be scattered of each mode in the interval $|k-k_L|
\leq \sigma _k/2$ is given by $|c_-|^2$. Then the probability of
scattering in the full beam is given by ${\cal N}_R |c_-|^2$, where
${\cal N}_R$ is the fraction of modes in the beam that can be
scattered:
\begin{equation}
{\cal N}_R = \int _{ k_L -\frac{1}{2}\sigma _k }^{ k_L +\frac{1}{2}\sigma _k }|f(k)|^2 dk = erf \left( \frac{\sigma _k}{\sqrt{2} \sigma} \right)
\end{equation}
with $erf$ the error function, $erf(\xi)=2\pi ^{-1/2}\int _0^{\xi} \exp (-u^2)du$.

On the other hand, the probability of transmission is the sum of two
contributions, (a) that of the modes outside the interval $|k-k_L|
\leq \sigma _k/2$, which cannot be scattered, and (b) another
corresponding to the probability of modes in the interval to be
transmitted without scattering, ${\cal N}_R|c_+|^2$. Adding both
contributions we have ${\cal N}_T + {\cal N}_R|c_+|^2$, where ${\cal
N}_T$ is the fraction of modes in the interval that cannot be
scattered. Clearly, we have ${\cal N}_T=erfc(\sigma
_k/(\sqrt{2}\sigma ))=1-{\cal N}_R$.

After the interaction we have two beams, one transmitted and the
other scattered, which we represent by the kets $|k_L>^{MM}$ and
$|-k_L>^{MM}$. As the overlapping between these beams is negligible
they can be considered orthonormal. Then although now we are in the
multi-mode regime, we can use a description for the longitudinal
variables with only two relevant kets because the detection process
can only discriminate between the alternatives represented by these
kets. If we would have used a mode-selective detection scheme the
description would be inadequate. The coefficients of the kets are
different from those associated with the single-mode case,
containing information about the multi-mode structure ($\sigma $)
and the effective window of scattering ($\sigma _k$). The expression
$c_+|k_L> + c_-|-k_L>$ must be replaced by $d_+|k_L>^{MM} +
d_-|-k_L>^{MM}$, with
\begin{equation}
d_+ =e^{-i\epsilon \tau}({\cal N}_T + {\cal N}_R |c_+|^2)^{1/2}  \; ; \; d_- ={\cal N}_R^{1/2} c_-
\end{equation}
where we have assumed that the relative phase between the reflected
and transmitted components is the same that in the case of
single-mode states (this is true for each mode). Clearly, we have
$|d_+|^2 +|d_-|^2=1$.

The relation between $\sigma $ and $\sigma _k$ gives the criterion
for the validity of the single-mode approximation. When $\sigma \ll
\sigma _k$, we have that $erf (\sigma _k/(\sqrt{2} \sigma))
\rightarrow 1$ and, consequently ${\cal N}_R \approx 1$ and ${\cal
N}_T \approx 0$. In this case, $d_{\pm} \approx c_{\pm} $ and it
makes sense to use the single-mode approximation.
\begin{figure}[H]
\center
\includegraphics[width=8.5cm,height=5.5cm]{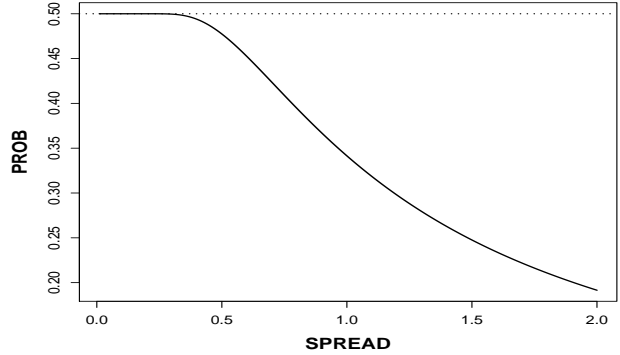}
\caption{Scattering probability $|d_-|^2$ versus the dimensionless
spread $\sigma /\sigma _k$ for $w=\pi /4$. The dotted line
represents the single-mode scattering probability $|c_-|^2=1/2$.}
\end{figure}
Concerning the {\it capital variables}, the state can be expressed
as $|K>_{G_{K_0}}^{MM}=\int dK G_{K_0}(K)|K>$. Thus, the complete
state can be expressed after the interaction as
\begin{equation}
|f(k,K)> = |K>_{G_{K_0}}^{MM} (d_+ |k_L>^{MM} + d_- |-k_L>^{MM})
\end{equation}
From this expression it follows that the probability of detecting,
for instance, a transmitted particle with transversal momentum $K$
is $|G_{K_0}(K)d_+|^2$. If we do not measure the perpendicular
momenta, as it is usually the case, the probability of having a
transmitted particle is the sum over $K$ of all these probabilities:
$\int |G_{K_0}(K)d_+|^2 dK=|d_+|^2$, because of the normalization
condition. This expression shows the same form obtained in the
single-mode case with the only change of $c_{\pm} \rightarrow
d_{\pm}$. This change can modify the dependence on $w$, in addition
to introduce the parameter $\tau$ (or the dimensionless spread $\sigma
_k/\sigma$).

We represent $|d_-|^2$ and $|c_-|^2$ in Fig. 4. For small values of
$\sigma /\sigma _k$ the scattering probabilities are very similar
for single- and multi-mode states. In contrast, when the value of
the ratio of spreads increases the multi-mode probability sharply
decreases with respect to the single-mode one.

\subsection{Two distinguishable particles}

In the case of two incident particles the central values of $g(k)$ are
equal (case $(I)$) or opposite (case $(II)$). On the other hand, for
$G$ we assume both widths to be equal ($\mu$) but the mean values
can be different ($G_{K_0}(K)$ and $G_{K_0 '}(K)$). The state of the
two-particle system can be expressed as $|f(k,K)>_1|f'  (k',K')>_2$,
with $f'(k,K)=g_{k_0'}(k)G_{K_0'}(K)$. As in the previous example we
assume that the final perpendicular momenta are not observed. Using
$ D_+ =e^{-i\epsilon \tau}({\cal M}_T + {\cal M}_R |c_+|^2)^{1/2}$
and $D_- ={\cal M}_R^{1/2} c_-$ with ${\cal M}_T + {\cal M}_R =1$,
for the coefficients of the second particle, we obtain
\begin{eqnarray}
{\cal P}_{MM}^{dis(I)}(k;k)= |d_+|^2|D_+|^2 \; ; \; {\cal P}_{MM}^{dis(I)}(-k;-k)= |d_-|^2|D_-|^2 \nonumber \\
{\cal P}_{MM} ^{dis(I)}(-k; k) =|d_-|^2|D_+|^2 \; ; \; {\cal P}_{MM} ^{dis(I)}(k;-k)= |d_+|^2|D_-|^2
\end{eqnarray}
and
\begin{eqnarray}
{\cal P}_{MM} ^{dis(II)}(-k; -k)= |d_-|^2|D_+|^2  \; ; \; {\cal P}_{MM} ^{dis(II)}(k;k)= |d_+|^2|D_-|^2   \nonumber \\
{\cal P}_{MM}^{dis(II)}(k;-k) = |d_+|^2|D_+|^2 \; ; \; {\cal
P}_{MM}^{dis(II)}(-k;k)=  |d_-|^2|D_-|^2
\end{eqnarray}
As $|d_+|^2 + |d_-|^2=1$ and $|D_+|^2 + |D_-|^2=1$ it is simple to
see that all these probabilities are correctly normalized. All the
probabilities are independent of the distributions $G_{K_0}$ as a
natural consequence of not observing the final momenta. The
dependence of these probabilities on $w$ clearly differs from that
on the case of single mode states. For instance, ${\cal
P}_{MM}^{dis(I)}(k;k)= {\cal N}_T{\cal M}_T + ({\cal N}_R {\cal M}_T
+ {\cal N}_T {\cal M}_R) \cos ^2 w + {\cal N}_R {\cal M}_R \cos ^4
w$. In addition we have the dependence on $\tau$ (or $\sigma _k/\sigma $). We shall later represent them in Fig. 5.

\subsection{Two identical particles}

The final state is $N(|f(k,K)>_1|f' (k',K')>_2 \pm
|f'(k,K)>_1|f(k',K')>_2)$. We have, for instance, $Nd_+D_+ |k_L>_1^{MM}
|k_L>_2^{MM}(|K>_{1,G_{K_0}}^{MM}|K'>_{2, G_{K'_0}}^{MM} \pm |K>_{1,
G_{K'_0}}^{MM}|K'>_{2, G_{K_0}}^{MM})$ for the two particles in the
channel $(k;k)$ in the case ($I$). The probability associated with the perpendicular
variables has the form $2\pm 2{\cal I}$, with
\begin{eqnarray}
{\cal I} = Re(_{G_{K_0},1}^{MM}<K|K>_{1,G_{K_0'}}^{MM} \; _{G_{K_0'},2}^{MM} <K'|K'>_{2,G_{K_0}}^{MM}) = \nonumber \\
\int dK \int dK' G_{K_0}(K) G_{K_0 '}(K')G_{K_0}(K')G_{K_0 '}(K)
\end{eqnarray}
The total probabilities become in the case $(I)$
\begin{eqnarray}
{\cal P}_{MM}^{(I)}(k;k)= 2N^2(1\pm {\cal I})|d_+|^2|D_+|^2  \; ; \; {\cal P}_{MM}^{(I)}(-k;-k)= 2N^2(1\pm {\cal I})|d_-|^2|D_-|^2 \nonumber \\
{\cal P}_{MM}^{(I)}(k;-k) \equiv {\cal P}_{MM}^{(I)}(k;-k) + {\cal
P}_{MM}^{(I)}(-k;k) = \nonumber \\
2N^2(|d_+|^2|D_-|^2+|D_+|^2|d_-|^2  \pm 2{\cal I}Re(d_+^* d_- D_+ D_-^*))
\end{eqnarray}
As usual, the normalization is obtained from the condition of the
sum of all the probabilities to be $1$, which reads $2N^2(1\pm {\cal
I}(|d_+||D_+|+|d_-||D_-|  )^2)=1$.

\begin{figure}[H]
\center
\includegraphics[width=10cm,height=7cm]{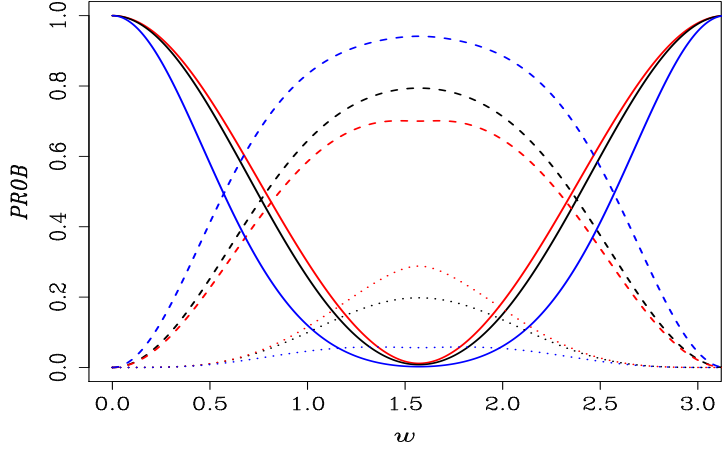}
\caption{As Fig. 2 for multi-mode states. Bosons, fermions and
distinguishable particles are represented by red, blue and black
lines. We represent the case $\mu =2$, $K_0=1$, $K_0 '=2$, ${\cal
N}_T=0.01$ and ${\cal M}_T=0.8$.}
\end{figure}
A specially simple situation is obtained when the two distributions
$g_{k_0} (k)$ are equal (this condition implies $\sigma = \sigma
'$), where we have $d_{\pm}=D_{\pm }$ and the normalization condition
is $N=1/\sqrt{2(1\pm {\cal I})}$. In this situation the probability
distributions for bosons, fermions and distinguishable particles are
equal. This result agrees with our previous discussion for
single-mode states. When the initial particles are in the same state
of the parallel variables the two alternatives
in the expression for the state of identical particles are actually
redundant and do not lead to distinctive exchange effects.
 
We represent the above results in figure 5. We consider the simpler
case, in which ${\cal N}_T$ and ${\cal N}_T$ are constant (we must
have a different spread of the multi-mode distribution for each $w$
and $\tau $). We take into account that ${\cal I}=\exp (-(K_0 - K_0
')^2/\mu ^2)$. At variance with the single-mode case we have that
the curves for bosons, and distinguishable particles (and now also
for fermions) are different. Moreover, the analytical form for the
case $(-k;-k)$ only shows a peak, whereas for the single-state there
were two separated ones. When the values of ${\cal N}_T$ and ${\cal
M}_T$ become close we recover the behavior observed for single-mode
states with curves almost similar in all the cases and two peaks for
$(-k;-k)$.
\begin{figure}[H]
\center
\includegraphics[width=10cm,height=7cm]{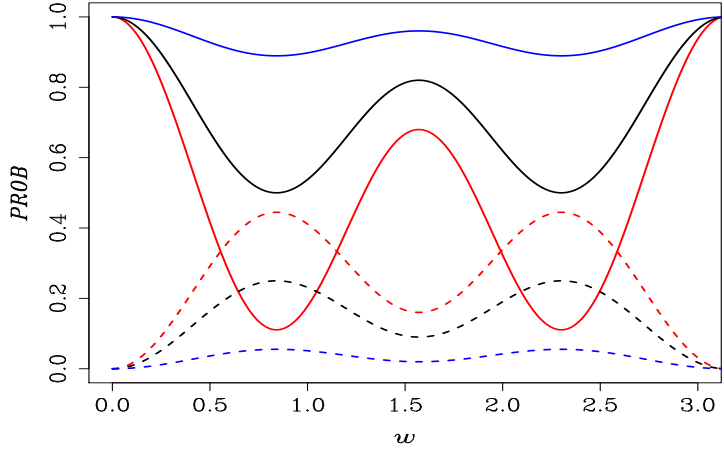}
\caption{The same as in Fig. 3 for multi-mode states with
$d_{\pm}=D_{\pm}$ and using the values $K_0=1$, $K_0'=2$, ${\cal
N}_T=0.1$ and $\mu =2$.}
\end{figure}

In the case $(II)$, in a similar way, we have
\begin{eqnarray}
{\cal P}_{MM}^{(II)}(k;-k)/ 2N^2 \equiv ({\cal P}_{MM}^{(II)}(k;-k)
+ {\cal P}_{MM}^{(II)}(-k;k))/ 2N^2= \nonumber \\
|d_+|^2|D_+|^2 + |d_-|^2|D_-|^2 \pm 2 {\cal I}Re(d_+^*D_- D_+^*d_- )
\\
{\cal P}_{MM}^{(II)}(k;k)/2N^2=(1\pm {\cal I}) |d_+|^2|D_-|^2  \; ;
\; {\cal P}_{MM}^{(II)}(-k; -k)/2N^2=(1 \pm {\cal I}) |d_-|^2|D_+|^2
\nonumber
\end{eqnarray}
The normalization condition is given by the expression $2N^2(1\pm {\cal I}(|d_+||D_-|-|d_-||D_+|)^2)=1$.

In the particular case $d_{\pm}=D_{\pm}$ ($N=1/\sqrt 2$) the above
expressions simplify to ${\cal P}_{MM}^{(II) eq }(k;-k)= |d_+|^4 +
|d_-|^4 \mp 2 {\cal I}|d_+|^2 |d_-|^2$, ${\cal P}_{MM}^{(II)
eq}(k;k)=(1\pm {\cal I}) |d_+|^2|d_-|^2$ and ${\cal P}_{MM}^{(II)
eq}(-k; -k)=(1 \pm {\cal I}) |d_-|^2|d_+|^2$. We represent them in
Fig. 6.

The curves resemble those obtained in the single-mode case. The most
important difference is that for fermions the possibility of
observing simultaneously two of them in the same output arm is not
null. This is due to the fact that now the perpendicular components
of the momentum are different, precluding the action of the
exclusion principle. It must also be noted that the visibility of
the bosonic $(k;-k)$ curve is slightly reduced. In the single-mode
scenario it was $1$, whereas now it does not reach that value
(${\cal P}_{max,MM}^{(II) eq}(k; -k)=1$, but ${\cal
P}_{min,MM}^{(II) eq}(k;-k) \neq 0$).

\section{Discussion: HOM-like experiments}

In this paper we have extended the theory of the two-particle
Kapitza-Dirac effect to the Bragg regime. With this extension we
complete the basic theory of the effect. We have derived the
detection probabilities for all the possible combinations of
scattering and transmission processes of the two particles. We have
developed a simple model based on some reasonable assumptions to
describe multi-mode states. Using this model we can quantify the
differences between single- and multi-mode states. In the case of
single-mode states all these probabilities can be expressed in terms
of the parameter $w$, whereas for multi-mode ones one also needs to
consider the duration of the interaction (or the dimensionless
spread). In the single-mode case we only have exchange effects for
equal perpendicular momenta, a restriction not present for
multi-mode states.

We have not discussed the possibility of carrying out experimental
tests of the results here derived. We refer to \cite{san} for a
brief presentation of the aspects related to the two-particle nature
of the arrangement, and \cite{Ba2,HBa} for the peculiarities
associated with the Bragg regime.

Bragg's scattering has been extensively studied for one-particle
systems. In the many-particle scenario one can also find in the
literature many analysis on the subject, mainly in the field of BEC
(see, for instance \cite{Pri}). However, there are still aspects of
the problem that deserve attention. In this paper we focus on the
aspects related to HOM-type experiments, which have played an
important role in quantum optics, triggering a lot of activity in
the field of two-photon interference experiments with beam
splitters. We explore if the same relevance could be expected for
massive systems, taking into account that the arrangement discussed
in this paper could be used as a two-particle beam splitter.

The first point to be noted is that the coefficients $c_+$ and $c_-$
play the same role of the transmission and reflection coefficients
in a beam splitter. The coefficients $c_{\pm}$ can be expressed in
terms of a single parameter $w$. In the optical case the
coefficients are complex variables that depend on the optical
frequency. In the massive case they are also complex, but depend on
the energy of the particle $e^{-i\epsilon \tau }$ and the potential
strength $V_0$ and the duration of the interaction $\tau $ via the
dimensionless parameter $w$.

Two interesting results emerge directly from our analysis.
The first result concerns to particles incident on the same arm of
the beam splitter. We have shown that when the particles are in
single-mode states or have the same multi-mode parallel distribution
there are not distinctive exchange effects and the behavior of
distinguishable and identical particles becomes equal. In quantum
optics we have a similar behavior. If two photons in the same state
are incident in the same input arm of the beam splitter, the
probabilities of finding the two photons in the different possible
combinations in the output arms are the same of two classical or
distinguishable particles (binomial distribution), without showing
any quantum interference effect \cite{Lou,Scu}. Our analysis gives
an intuitive explanation for the absence of distinctive exchange
effects, only based on the physical meaning of the different terms
of the state vector. At variance with \cite{Lou,Scu}, we have
demonstrated that exchange effects can be present in the multi-mode
case, giving rise to some differences between distinguishable and
identical particles. Up to our knowledge, this behavior has not been
previously described in the literature. Modifying the multi-mode
distributions we can modulate the differences between
distinguishable and identical particles. This is an unexplored
scenario where some new physical effects could emerge.

Our second result refers to the presence of dips in the case $(II)$.
In quantum optics the HOM dip takes place for a perfect temporal
overlapping of the two photons arriving on different arms of the
beam splitter: the two photons are always found in the same output
arm. In the massive case, Fig. 3 shows dips in the boson curve for
some values of $w$. At these values there are not coincidence
detections in the two exit paths. Thus, one of the most
characteristic signatures of HOM interferometry is also present in
the bosonic massive case, reinforcing the analogy between massless
and massive bosons. Note a difference between the massless and
massive cases. In the first one the parameters of the beam splitter
(transmissivity and reflectivity) are fixed and the temporal
overlapping between the photons varies. In the second one, the
perfect overlapping between the two particles is assumed and one
must vary $w$, the {\it beam splitter parameter}.

Finally, we shall propose a potential application of the
two-particle massive beam splitter, a scheme for the verification of
(anti)symmetrization. The HOM arrangement was originally conceived
for precision measurements of time intervals between the arrivals of
photons. Similarly, we could use our arrangement to determine if the
overlapping between the wave functions of the two identical
particles is large or not. If one wants to prepare identical
particles in (anti)symmetrized states for some physical task, one
must have some method to test that the particles are actually in
that state. This can be done via double Bragg's scattering. When the
overlapping is large, the two wave functions must be
(anti)symmetrized and the results derived in the previous sections
for identical particles hold. In contrast, with a lower degree of
overlapping the behavior of the particles becomes increasingly
similar to that of distinguishable particles. These properties can
be used to quantitatively measuring the degree of overlapping
between the two identical particles.

{\bf Acknowledgments} We acknowledge partial support from MEC (CGL 2007-60797).

\end{document}